\documentclass[aps,prd,reprint,superscriptaddress,preprintnumbers]{revtex4-1}
\usepackage{amssymb,amsfonts,amsmath,graphicx,hyperref}
\usepackage[utf8]{inputenc}
\usepackage{bm}
\usepackage{bbold}
\usepackage{enumitem}
\usepackage{color}

\allowdisplaybreaks

\def\eqn#1{eq.~(\ref{#1})}
\def\Eqn#1{Equation~(\ref{#1})}
\def\eqns#1#2{eqs.~(\ref{#1}) and~(\ref{#2})}

\def\rcite#1{ref.~\cite{#1}}
\def\rcites#1{refs.~\cite{#1}}

\def\bra#1{\langle #1|}
\def\ket#1{|#1 \rangle}

\def\braket#1{\langle #1 \rangle}

\def\ie{{\it i.e.} }
\def\eg{{\it e.g.}}

\def\cO{{\cal O}}

\def\be{\begin{equation}}
\def\ee{\end{equation}}
\def\bea{\begin{eqnarray}}
\def\eea{\end{eqnarray}}  
\def\beal{\begin{equation}\begin{aligned}}
\def\eeal{\end{aligned}\end{equation}}
\def\nn{\nonumber}
\def\braket#1{\langle #1 \rangle}

\def\vs{\varsigma}

\begin{document} 

\preprint{
UUITP–40/23,
NORDITA 2023-117 
}

\title{Compton Amplitude for Rotating Black Hole from QFT}

\author{Lucile Cangemi}
\email{lucile.cangemi@physics.uu.se}
\affiliation{Department of Physics and Astronomy, Uppsala University, Box 516, 75120 Uppsala, Sweden,}

\author{Marco Chiodaroli}
\email{marco.chiodaroli@physics.uu.se}
\affiliation{Department of Physics and Astronomy, Uppsala University, Box 516, 75120 Uppsala, Sweden,}

\author{Henrik Johansson}
\email{henrik.johansson@physics.uu.se}
\affiliation{Department of Physics and Astronomy, Uppsala University, Box 516, 75120 Uppsala, Sweden,}
\affiliation{Nordita, Stockholm University and KTH Royal Institute of Technology, Hannes Alfv\'{e}ns  v\"{a}g 12, 10691 Stockholm, Sweden,}

\author{Alexander Ochirov}
\email{ochirov@shanghaitech.edu.cn}
\affiliation{School of Physical Science and Technology, ShanghaiTech University, 393 Middle Huaxia Road, Shanghai 201210, China,}
\affiliation{London Institute for Mathematical Sciences, Royal Institution,
21 Albemarle St, London W1S 4BS, UK,}

\author{Paolo Pichini}
\email{p.pichini@qmul.ac.uk}
\affiliation{Centre for Theoretical Physics, Department of Physics and Astronomy,
Queen Mary University of London, Mile End Road, London E1 4NS, UK,}
\affiliation{Department of Physics and Astronomy, Uppsala University, Box 516, 75120 Uppsala, Sweden,}

\author{Evgeny Skvortsov}
\email{evgeny.skvortsov@umons.ac.be}
\affiliation{Service de Physique de l’Univers, Champs et Gravitation,
Universit\'{e} de Mons, 20 place du Parc, 7000 Mons, Belgium,}
\affiliation{Lebedev Institute of Physics, Leninsky avenue 53, 119991 Moscow, Russia}

\begin{abstract}
We construct a candidate tree-level gravitational Compton amplitude for a rotating Kerr black hole, for any quantum spin $s=0,1/2,1,\dots,\infty$, from which we extract the corresponding classical amplitude to all orders in the spin vector $S^\mu$. We use multiple insights from massive higher-spin quantum field theory, such as massive gauge invariance and improved behavior in the massless limit. A chiral-field approach is particularly helpful in ensuring correct degrees of freedom, and for writing down compact off-shell interactions for general spin. The simplicity of the interactions is echoed in the structure of the spin-$s$ Compton amplitude, for which we use homogeneous symmetric polynomials of the spin variables. Where possible, we compare to the general-relativity results in the literature, available up to eighth order in spin.
\end{abstract}


\maketitle

%%%%%%%%%%%%%%%%%%%%%%%%%%%%%%%%%%%%%%%%%%%%%%%%%%%%
\section{Introduction}
\label{sec:intro}
%%%%%%%%%%%%%%%%%%%%%%%%%%%%%%%%%%%%%%%%%%%%%%%%%%%%

Gravitational dynamics of extended rotating objects in general relativity (GR) is a difficult computational problem. Approaches include numerical relativity, and relevant to the current work, simplifying analytic limits, such as point-like approximations, dressed with an infinite set of effective couplings that control multipole moments, tidal parameters and other degrees of freedom.
Black holes (BHs) are special in that they are known to possess a remarkable simplicity, which may transpire to making computations easier.
The simplicity is rooted in the familiar no-hair theorem, which states that a macroscopic black hole  may be characterized entirely by its mass, angular momentum, and  electric charge, where the latter can be neglected for astrophysical black holes.

The gravitational dynamics of a rotating BH must thus be parametrized by its mass~$m$ (or momentum $p^\mu$ in a generic frame) and spin $S^\mu = m a^\mu$, where the latter variable is a transverse vector $p\cdot a=0$ with units of length, corresponding to the Kerr-metric~\cite{Kerr:1963ud} ring singularity of radius $|a|:=\sqrt{-a^2}$.
In the effective worldline approach~\cite{Goldberger:2004jt,Porto:2005ac,Levi:2015msa,Porto:2016pyg,Levi:2018nxp} (see also recent work~\cite{Jakobsen:2021lvp,Jakobsen:2021zvh,Saketh:2022wap,Jakobsen:2023ndj, Ben-Shahar:2023djm,Scheopner:2023rzp}), the BH spin-multipole interactions linear in the Riemann tensor have simple combinatorial coupling coefficients.
These coefficients originate from the exponential structure of these couplings~\cite{Vines:2017hyw} due to the Newman-Janis shift relationship~\cite{Newman:1965tw} between the Kerr and Schwarzschild solutions.

Scattering amplitudes have recently emerged as an efficient way to encode black-hole dynamics~\cite{Damour:2017zjx,Bjerrum-Bohr:2018xdl,Cheung:2018wkq,Guevara:2018wpp,Chung:2018kqs,
Bern:2019nnu,Bern:2019crd,Aoude:2020onz,Chung:2020rrz,
Bern:2020buy,Kosmopoulos:2021zoq,
DiVecchia:2021bdo,Bjerrum-Bohr:2021din,Brandhuber:2021eyq,Chen:2021kxt,Bern:2021yeh,Alessio:2022kwv,
Aoude:2022trd,Bern:2022kto,Buonanno:2022pgc,Aoude:2022thd,FebresCordero:2022jts,Menezes:2022tcs,Bjerrum-Bohr:2023jau,
Haddad:2023ylx,Herderschee:2023fxh,Elkhidir:2023dco,Bautista:2023szu,
Aoude:2023vdk,Damgaard:2023ttc,Aoude:2023fdm,Bern:2023ity,Bjerrum-Bohr:2023iey,Bini:2023fiz,DeAngelis:2023lvf,Brandhuber:2023hhl,Aoude:2023dui,Chen:2023qzo,Georgoudis:2023eke, Luna:2023uwd}.
The classical three-point Kerr amplitude, which exists for analytically continued momenta, has a simple exponential form
${\cal M}(\bm{1},\bm{2},3^\pm) = {\cal M}_3^{(0)} e^{\pm p_3 \cdot a}$
\cite{Guevara:2018wpp,Chung:2018kqs,Guevara:2019fsj,Arkani-Hamed:2019ymq,Guevara:2020xjx},
where ${\cal M}_3^{(0)}=-\sqrt{32\pi G_\text{N}} (p_1 \cdot \varepsilon_3^\pm)^2$ is the corresponding amplitude for a Schwarzschild black hole of momentum $p=p_1$, and graviton with momentum~$p_3$ and helicity~$\pm 2$.
The quantum counterpart of the three-point Kerr amplitude was known even earlier, as it was first written down by Arkani-Hamed, Huang and Huang~\cite{Arkani-Hamed:2017jhn} for gravitationally interacting massive spin-$s$ particles,
\begin{equation}
\label{3pt}
{\cal M}(\bm{1}^s\!,\bm{2}^s\!,3^+) = {\cal M}_3^{(0)} \frac{\braket{\bm{12}}^{2s}}{m^{2s}} 
~\xrightarrow[s \to \infty]{\;\hbar s \to m|a|\;}~ {\cal M}_3^{(0)} e^{p_3 \cdot a},
\end{equation}
where $\braket{\bm{12}}$ is a spinor product of two chiral Weyl spinors $\ket{\bm{i}}$ satisfying the Dirac equation, see notation~\footnote{We use the massive spinor-helicity formalism of \rcite{Arkani-Hamed:2017jhn} (for earlier iterations, see \rcites{Kleiss:1986qc,Dittmaier:1998nn,Schwinn:2005pi,Conde:2016vxs,Conde:2016izb}).
The angle and square brakets denote Weyl 2-spinors corresponding to the indicated momentum:
$p_{i\mu} \sigma^\mu_{\alpha\dot{\beta}} = \ket{i^a}_\alpha \epsilon_{ab} [i^b|_{\dot{\beta}}$.
Their Lorentz and little-group indices may be contracted with identical ${\rm SL}(2,\mathbb{C})$ or SU(2) Levi-Civita symbols, respectively, such that $\epsilon^{12} = 1$.
The spinor products are
$\braket{1^a 2^b}:=\epsilon^{\beta\alpha} \ket{1^a}_\alpha \ket{2^b}_\beta$
and
$[1^a 2^b]:=\epsilon^{\dot{\alpha}\dot{\beta}}[1^a|_{\dot{\alpha}} [2^b|_{\dot{\beta}}$.
The SU(2) indices $a,b$ can be absorbed by contracting with complex wavefunctions $z_i^a$, as indicated by the bold notation $|\bm i] = |i^a] z_{i a}$ and $\ket{\bm i} = \ket{i^a} z_{i a}$, see \rcite{Chiodaroli:2021eug}.
The four-dimensional Pauli matrices relating tensors and bispinors are
$\sigma^\mu_{\alpha\dot{\beta}} := (1,\vec{\sigma})$ and
$\bar{\sigma}^{\mu,\dot{\alpha}\beta} := \epsilon^{\dot{\alpha}\dot{\gamma}} \epsilon^{\beta\delta} \sigma^\mu_{\delta\dot{\gamma}}
= (1,-\vec{\sigma})$.
Our amplitudes use momenta that are all incoming.
}.

In this letter, we push to the next level the idea that quantum field theory can non-trivially inform us of classical Kerr amplitudes. We use recent higher-spin advances~\cite{Zinoviev:2001dt,Zinoviev:2006im,Zinoviev:2008ck,Zinoviev:2009hu,Ochirov:2022nqz,Cangemi:2022bew,Cangemi:2022abk,Cangemi:2023ysz} and construct a plausible four-point Compton amplitude describing tree-level dynamics of a rotating Kerr black hole to all orders in spin.
The tree level means that we work to lowest order in the gravitational constant $G_\text{N}$, and thus we are not sensitive to non-perturbative effects near the BH horizon.
We will provide evidence in favor of our proposal at lower orders in spin, comparing to some of the recent results~\cite{Chia:2020yla,Bautista:2021wfy,Ivanov:2022qqt,Bautista:2022wjf,Saketh:2023bul,Bautista:2023sdf} from solving the Teukolsky equation from black-hole perturbation theory (BHPT)~\cite{Teukolsky:1973ha,Press:1973zz,Teukolsky:1974yv}.

We proceed by introducing the needed higher-spin theory framework, leading to an explicit chiral-field Lagrangian.
Based on the cubic higher-spin  interactions, we obtain a quantum Compton amplitude for any spin, which we supplement with suitable contact terms built out of symmetric homogeneous polynomials.
In the classical limit, the Compton amplitude becomes an entire function of the ring-radius vector, and we discuss its properties.

%%%%%%%%%%%%%%%%%%%%%%%%%%%%%%%%%%%%%%%%%%%%%%%%%%%%
\section{Massive higher-spin theory}
\label{sec:HigherSpin}
%%%%%%%%%%%%%%%%%%%%%%%%%%%%%%%%%%%%%%%%%%%%%%%%%%%%

Rotating black holes have astronomically large spin in natural units, and any effective point-like QFT description of such objects necessarily involves a higher-spin theory framework. However, constructing theories of massive higher-spin particles~\cite{Fierz:1939ix,Singh:1974qz,Singh:1974rc} is generally considered to be a formidable endeavor.
Early work discussed a range of pathologies in such theories (\eg~\rcites{Johnson:1960vt,Velo:1969bt}), but more recently the problems have been understood to arise from the proliferation of unphysical degrees of freedom.
Composite particles with spins larger than two do, of course, exist in nature and may thus be described by effective field theories (EFTs) with various ranges of validity.
In particular, the theoretically allowed range of BH masses suggests that Kerr EFTs should possess a relatively wide range of validity~\cite{Cangemi:2022bew}.
We are therefore led to believe that the BH simplicity also implies enhanced properties, when it comes to derivative power counting and tree-level unitarity~\cite{Ferrara:1992yc,Porrati:1993in,Cucchieri:1994tx,Chiodaroli:2021eug} of their EFTs.

As for the higher-spin challenge of the unphysical degrees of freedom, the following two complementary approaches have been studied:

{\bf Massive higher-spin gauge symmetry.} As formalized by Zinoviev~\cite{Zinoviev:2001dt,Zinoviev:2006im,Zinoviev:2008ck,Zinoviev:2009hu,Zinoviev:2010cr,Buchbinder:2012iz}, massive higher-spin gauge symmetry is a natural generalization of the concept familiar from spontaneously broken Yang-Mills theory, as well as from the St\"uckelberg mechanism (see \eg~section~2 of \rcite{Cangemi:2023ysz}).
The unphysical longitudinal modes in the conventional traceless symmetric tensor field~$\Phi_{\mu_1\dots\mu_s}$ are compensated by a tower of auxiliary degrees of freedom, so that a single spin-$s$ particle is described by in total $s{+}1$ double-traceless symmetric tensors $\Phi^{\mu_1\dots\mu_k} =: \Phi^k$ of rank~$k=0,\ldots,s$.
The trick~\cite{Zinoviev:2001dt} is to endow this field content with a local gauge symmetry.
This symmetry must be suitably adjusted and enforced perturbatively in accord with the interactions.
This then guarantees the consistency of the resulting interacting higher-spin theory at the cost of increasingly complicated symmetry structure with each order in the coupling constant.
In \rcite{Cangemi:2022bew} we showed that along with certain assumptions,
including the maximum of $2s-2$ derivatives in the three-point vertex, 
the massive gauge symmetry constrains the three-point amplitude to the form~\eqref{3pt}, see also \rcite{Skvortsov:2023jbn} for a cubic action.

{\bf Chiral fields.} The novel chiral approach~\cite{Ochirov:2022nqz} to massive higher spins in four dimensions entirely sidesteps the issue of unphysical modes by switching to the symmetric spinor field $\Phi_{\alpha_1\ldots \alpha_{2s}} =: \ket{\Phi}$ in the chiral $(2s,0)$ representation of the Lorentz group ${\rm SL}(2,\mathbb{C})$.
The price to pay is that of obscuring parity, which at the level of amplitudes largely amounts to swapping chiral and antichiral spinors (denoted by angle and square brackets, respectively).
At the Lagrangian level this switch is non-trivial, because we commit to the massive fields being strictly chiral.

Ref.~\cite{Ochirov:2022nqz} considered the simplest chiral action
$\tfrac{1}{2}\!\int\!\!d^4 x \sqrt{-g} \big( \braket{\nabla_\mu \Phi| \nabla^\mu \Phi} {-} m^2 \braket{\Phi|\Phi} \big)$, with $\bra{\Phi}\!:=\!\Phi^{\alpha_1\ldots \alpha_{2s}}$
and the covariant derivatives involve the spin connection:
\beal
\nabla_\mu \Phi_{\alpha_1\dots\alpha_{2s}} = &\,
   \partial_\mu \Phi_{\alpha_1\dots\alpha_{2s}}
 + 2s\,\omega_{\mu,(\alpha_1}{}^\beta \Phi_{\alpha_2\dots\alpha_{2s})\beta} , \\
   \omega_{\mu,\alpha}{}^\beta := &\,
   \frac{1}{4} \omega_\mu^{\hat{\nu}\hat{\rho}} 
   \sigma_{\hat{\nu},\alpha\dot{\gamma}}
   \bar{\sigma}_{\hat{\rho}}^{\dot{\gamma}\beta} ,
\label{CovDerivativeG}
\eeal
where the frame indices are marked with hats.
The simple action generates a positive-helicity $n$-point amplitude
\begin{equation}
\label{AllPlus}\!\!
{\cal M}(\bm{1}^s\!,\bm{2}^s\!,3^+\!,\dots,n^+) = \frac{\braket{\bm{12}}^{2s}\!}{m^{2s}} {\cal M}(\bm{1}^0\!,\bm{2}^0\!,3^+\!,\dots,n^+) ,
\end{equation}
whose factorization channels self-consistently produce lower-point Kerr amplitudes~\cite{Johansson:2019dnu,Aoude:2020onz,Lazopoulos:2021mna}, including the established three-point amplitude~\eqref{3pt}.
When it comes to negative-helicity amplitudes, however, it fails to reproduce known parity conjugates, starting with the three-point amplitude
\begin{equation}\!\!
\label{3ptConj}
{\cal M}(\bm{1}^s\!,\bm{2}^s\!,3^-) = {\cal M}_3^{(0)} \frac{[\bm{12}]^{2s}}{m^{2s}} .
\end{equation}
Fixing this, and thus restoring parity at the three-point level, requires adding the following Lagrangian term
\begin{equation}\!\!\!\!
\begin{aligned} \label{eq:longformLag}
-\frac{1}{4}\!\int\!d^4x \sqrt{-g} \sum_{k=0}^{2s-2} \frac{2s{-}k{-}1}{m^{2k}}
  (\nabla_{\alpha_1\dot{\gamma}_1}\!\cdots\!\nabla_{\alpha_k\dot{\gamma}_k}
  \Phi^{\alpha_1\dots\alpha_{2s}}) & \\ \times
  R_{-\alpha_{k+1}}{}^{\beta_{k+1}}{}_{\alpha_{k+2}}{}^{\beta_{k+2}}
  \delta_{\alpha_{k+3}}^{\beta_{k+3}}\!\cdots
  \delta_{\alpha_{2s}}^{\beta_{2s}} & \\ \times
  (\nabla^{\dot{\gamma}_1\beta_1}\!\cdots\!\nabla^{\dot{\gamma}_k\beta_k}
  \Phi_{\beta_1\dots\beta_{2s}}) & .
\end{aligned}\!\!\!
\end{equation}
Here the chiral Riemann or Weyl curvature spinor is
\begin{equation}
\label{Rminus}
R_{-\alpha}{}^\beta{}_\gamma{}^\delta := \frac{1}{4}
  R_{\hat{\lambda}\hat{\mu}\,\hat{\nu}\hat{\rho}}
  \sigma^{\hat{\lambda}}_{\alpha\dot{\varepsilon}}
  \bar{\sigma}^{\hat{\mu},\dot{\varepsilon}\beta}
  \sigma^{\hat{\nu}}_{\gamma\dot{\zeta}}
  \bar{\sigma}^{\hat{\rho},\dot{\zeta}\delta} .
\end{equation}
A shorter way to write the resulting Lagrangian is provided by the ${\rm SL}(2,\mathbb{C})$-bracket notation:
\begin{widetext}
\begin{equation}
\label{ActionGravAHH}
{\cal L}_\text{Kerr}
= \sqrt{-g} \bigg\{ \frac{1}{2} \braket{\nabla_\mu \Phi|\nabla^\mu \Phi}
- \frac{m^2\!}{2} \braket{\Phi|\Phi}
- \frac{1}{4} \sum_{k=0}^{2s-2} \frac{2s{-}k{-}1}{m^{2k}}
  \bra{\Phi} \Big\{
  \big(|\overset{\leftarrow}{\nabla}|\overset{\rightarrow}{\nabla}|\big)^{\odot k}\!\odot |R_-|
\Big\} \ket{\Phi} \bigg\}
+ {\cal O}(R^2) .
\end{equation}
\end{widetext}
The arrows over the derivatives indicate which matter field the derivative is acting on, while the $\odot$ sign denotes the symmetrized tensor product.

Checking that the action~\eqref{ActionGravAHH} generates the Kerr three-point amplitudes~\eqref{3pt} and~\eqref{3ptConj} is straightforward and analogous to the gauge-theory case, where a root-Kerr Lagrangian takes the form~\cite{Cangemi:2023ysz}
\begin{align}
\label{ActionGaugeAHH}
{\cal L}_{\sqrt{\text{Kerr}}} &
= \braket{D_\mu \Phi|D^\mu \Phi} - m^2 \braket{\Phi|\Phi} \\* &
 + \sum_{k=0}^{2s-1} \frac{ig}{m^{2k}} \bra{\Phi}
   \Big\{ |\overset{\leftarrow}{D}|\overset{\rightarrow}{D}|^{\odot k}\!
          \odot |F_-| \Big\} \ket{\Phi} + {\cal O}(F^2) . \nn
\end{align}
Here $D_\mu = \partial_\mu - ig A_\mu$, and $|F_-|$ stands for the Maxwell spinor
$F_{-\alpha}{}^\beta := \tfrac{1}{2} F_{\mu\nu} \sigma^\mu_{\alpha\dot{\gamma}} \bar{\sigma}^{\nu,\dot{\gamma}\beta}$.
In \rcite{Cangemi:2023ysz}, this Lagrangian was shown to give the amplitudes with an identical massive-spin structure to \eqns{3pt}{3ptConj}.
Note that the explicit spin dependence of the latter action is contained exclusively as the highest power of derivatives in the three-point vertex.
The form of the vertex belongs to a family of complete homogeneous symmetric polynomials of $n$ variables $\{\vs_1,\dots,\vs_n\}$ \cite{Cangemi:2023ysz},
\begin{equation}
\label{PolyDef}
P_n^{(k)} := \sum_{i=1}^n \frac{\vs_i^k}{\prod_{j \neq i}^n (\vs_i-\vs_j)} = \sum_{\sum l_i=k-n+1}\!\vs_1^{l_1}\!\cdots \vs_n^{l_n}\,.
\end{equation}
Namely, the linear-in-$|F_-|$ term in the Lagrangian~\eqref{ActionGaugeAHH} is a geometric sum with derivative structure identical to $P_2^{(2s)}\big(1,|\overset{\leftarrow}{D}|\overset{\rightarrow}{D}|/m^2\big)$, with ordinary products replaced by tensor products.
It is thus perhaps not surprising that the spin dependence of the resulting four-point (Compton) amplitude was also simply expressed in \rcites{Cangemi:2022bew,Cangemi:2023ysz} in terms of such polynomials, namely
$P_1^{(2s)}$, $P_2^{(2s)}$, $P_2^{(2s-1)}$, $P_4^{(2s)}$ and $P_4^{(2s-1)}$.
In this paper, we confirm that this structural property generalizes to the gravitational massive higher-spin theory, and gravitational Kerr Compton amplitude, including the contact terms undetermined by \eqn{ActionGravAHH}.

Although the gravitational action~\eqref{ActionGravAHH} looks very similar to the gauge-theory one, we note that the linear-in-$|R_-|$ interactions explicitly grow with the spin quantum number~$s$.
Remarkably, the $s$ dependence and derivative structure is exactly reproduced by the three-variable polynomial $P_3^{(2s)}\big(1,|\overset{\leftarrow}{\nabla}|\overset{\rightarrow}{\nabla}|/m^2,1\big)$ with a repeated argument.
In the following, we will present our proposal for the gravitational Compton amplitude featuring similar polynomials with repeated arguments.

%%%%%%%%%%%%%%%%%%%%%%%%%%%%%%%%%%%%%%%%%%%%%%%%%%%%
\section{Gravitational Compton amplitudes}
\label{sec:Compton}
%%%%%%%%%%%%%%%%%%%%%%%%%%%%%%%%%%%%%%%%%%%%%%%%%%%%

We are interested in the general spin-$s$ gravitational Compton amplitude for a rotating Kerr black hole, at tree level $G_\text{N}\, m \ll |a|$. A candidate quantum amplitude with correct factorization properties (inherited from our cubic Lagrangian \eqref{ActionGravAHH}) is given by ${\cal M}_4=8 \pi G_\text{N} M_4$, where
\begin{widetext}
\begin{align}
\label{eq:quantumcompton}
M(\bm{1}^s\!,\bm{2}^s\!,3^-\!,4^+)  = & \, \frac{\langle3|1|4]^4  P_1^{(2s)}}{m^{4s} s_{12}t_{13}t_{14}}
-\frac{\braket{\bm{1}3} [{4\bm 2}] \langle3|1|4]^3 }{ m^{4s} s_{12}t_{13}} P_2^{(2s)}
+\frac{\braket{\bm{1}3} \braket{3{\bf 2}} [\bm{1}4] [4\bm{2}]}{m^{4s} 
 s_{12}} \big(\langle3|1|4]^2 P_2^{(2s-1)}\!+ 
 m^4 \langle3|\rho|4]^2 P_4^{(2s-1)} \big) \nn \\ &
 +\frac{\braket{\bm{1}3} \braket{3{\bf 2}} [\bm{1}4] [4\bm{2}]}{m^{4s-2} s_{12}} \langle3|1|4] \langle3|\rho|4] \big(P_2^{(2s-2)} - m^2\braket{\bm{12}}[\bm{12}] P_4^{(2s-2)} \big) \\ &
 + \frac{\braket{\bm{1}3}^2 \braket{3{\bf 2}}^2 [\bm{1}4]^2 [4\bm{2}]^2}{2m^{4s-4} }\braket{\bm{12}}[\bm{12}]
   \Big[(1+\eta) P_{5|\vs_1}^{(2s-2)} + (1-\eta) P_{5|\vs_2}^{(2s-2)} \Big]
 + \alpha\, C^{(s)}_\alpha . \nn
\end{align} 
\end{widetext}
We now explain this formula. The symmetric polynomials $P_n^{(k)}$ were defined in \eqn{PolyDef},
but now we globally identify the $\vs_i$ with four spin-dependent variables:
\beal
\label{VarSigmaDef}
\vs_1 & := \bra{\bm{1}}4|\bm{2}] - m[\bm{12}] , \qquad~\;\,\quad
\vs_3 := -m \braket{\bm{12}} , \\*
\vs_2 & := - \bra{\bm{2}}4|\bm{1}] - m [\bm{12}] , \qquad \quad
\vs_4 := -m [\bm{12}] .
\eeal
We also need polynomials where a variable is inserted twice, abbreviated as $P^{(k)}_{n|\vs_i} := \lim_{\vs_n \rightarrow \vs_i} P_n^{(k)}$,
and further repetitions are  
$P^{(k)}_{n|\vs_i\vs_j} := \lim_{\vs_{n-1} \rightarrow \vs_j} P^{(k)}_{n|\vs_i}$, etc.
The repeated variables are still universally identified with those in \eqn{VarSigmaDef}.
For brevity, we have also expressed the Compton amplitude using $\bra{3}\rho|4] := \braket{3\bm{1}} [\bm{2}4] + \braket{3\bm{2}}[\bm{1}4] $, where the $\rho^\mu$ vector was introduced in \rcite{Cangemi:2023ysz}. 
We postpone the discussion of $\alpha$ and $\eta$ (notation inherited from~\rcite{Bautista:2022wjf}), which are convenient tags for certain contact terms, corresponding to the unknown ${\cal O}(R^2)$ terms in \eqn{ActionGravAHH}.

In order to fix the {\it a priori} unknown contact terms, given on the last line of \eqn{eq:quantumcompton}, we impose the following constraints on the amplitude:
\begin{enumerate}
\item[(i)] well behaved $s \to \infty$ limit;
\item[(ii)] improved behavior in $m \to 0$ limit: finite for $s\leq 2$ and otherwise $M(\bm{1}^s\!,\bm{2}^s\!,3^-\!,4^+) \sim m^{-4s+4}$, see \footnote{The imposed $m\to0$ behavior is equivalent to the good high-energy behavior of cubic amplitudes~\cite{Arkani-Hamed:2017jhn}, which is easiest to see with non-chiral higher-spin fields~\cite{Cangemi:2022bew, Cangemi:2023ysz}.};
\item[(iii)] contact terms proportional to the structure $\braket{\bm{1}3}^2 \braket{3\bm{2}}^2 [\bm{1}4]^2 [4\bm{2}]^2$;
\item[(iv)] all helicity-independent factors written in terms of the symmetric polynomials~$P_n^{(k)}$;
\item[(v)] parity invariance;
\item[(vi)] the classical-spin hexadecapole $S^4$ is fixed by the established~\cite{Siemonsen:2019dsu} $s=2$ amplitude~\cite{Arkani-Hamed:2017jhn}.
\end{enumerate}

First, we consider the simplest contact-term solution of this form, and it is given by the last line of \eqn{eq:quantumcompton}, evaluated at $\alpha = 0 = \eta$.
The amplitude~\eqref{eq:quantumcompton} reproduces the known Compton amplitudes for massive particles with spin $s\leq 2$, first discussed in \rcite{Arkani-Hamed:2017jhn}.
Moreover, it matches the $s = 5/2$ amplitude proposed in \rcite{Chiodaroli:2021eug}.
In fact, the last line, containing two auxiliary parameters~$\eta$ and $\alpha$, is identically zero for $s\le 5/2$ as a consequence of the imposed constraints.
Next, we choose to include the contact terms proportional to the parameter~$\eta$ (to be described in more detail below), such that they match the corresponding dissipative terms appearing in the classical 32-pole $S^5$ of \rcite{Bautista:2022wjf}.
The second auxiliary parameter~$\alpha$ captures deviations away from our preferred contact-term solution, and it coincides with the same parameter of \rcite{Bautista:2022wjf}. Finally, we note that other simple contact candidates exist with similar properties, such as in footnote~\footnote{To obtain an alternative contact term, swap out $P_{5|\vs_i}^{(2s-2)} \rightarrow \tfrac{1}{\varsigma_3\varsigma_4}\big(P_{5|\vs_i}^{(2s)} {-}  P_4^{(2s-1)} {-} P_{3|\vs_i}^{(2s-2)}{+}P_2^{(2s-3)})$ in \eqn{eq:quantumcompton}. The classical limit is unchanged; out of the constraints (i)--(vi) only the $m$-scaling is relaxed to $m^{-4s-3}$.}.

In the following sections, we will see that the classical limit of the full quantum Compton amplitude \eqref{eq:quantumcompton} reproduces results derived in black-hole perturbation theory (BPHT)
in \rcites{Bautista:2022wjf,private}, including dissipative contributions.
Moreover, we will extract novel predictions to all orders in spin from the $\alpha = 0$ part.

%%%%%%%%%%%%%%%%%%%%%%%%%%%%%%%%%%%%%%%%%%%%%%%%%%%%
\section{Classical limit of gravity amplitudes}
\label{sec:ClassicalLimit}
%%%%%%%%%%%%%%%%%%%%%%%%%%%%%%%%%%%%%%%%%%%%%%%%%%%%

To study the classical limit of \eqn{eq:quantumcompton}, we use the approach of \rcite{Aoude:2021oqj}, which is based on coherent spin states (see \eg~\rcites{Atkins:1971zy,Radcliffe_1971,Perelomov_1977}), which was shown in~\rcite{Cangemi:2023ysz} to be highly efficient when combined with the homogeneous symmetric polynomials~\eqref{PolyDef}.

We begin by defining appropriate kinematic variables and their scaling in the $\hbar \to 0$ limit~\cite{Kosower:2018adc}:
\beal
\label{eq:clMomScaling}
p^\mu & := p_1^\mu \sim 1 , \qquad \qquad~\:\quad
q^\mu := (p_3 + p_4)^\mu \sim \hbar , \\
q_\perp^\mu & := (p_4 - p_3)^\mu \sim \hbar , \qquad
\chi^\mu := \bra{3}\sigma^\mu|4] \sim \hbar ,  
\eeal
where Lorentz invariants scale as
\beal
\label{ClassicalLimitScaling1}
& p^2 = m^2 \sim 1 , \qquad \quad
q_\perp^2 = 2 p \cdot q = -q^2 \sim \hbar^2 , \\
& p \cdot q_\perp \sim \hbar , \qquad\:\qquad
p \cdot \chi \sim \hbar .
\eeal
Next, we study the spin degrees of freedom of the massive particles.
Particles~1 and~2 describe an incoming and outgoing black-hole state, respectively.
We choose $p_1$ to be the default black-hole four-momentum and expand the spinors for particle~2 via a boost of particle~1:
\begin{equation}
\label{eq:boost}
\begin{aligned}
|\bm{2}\rangle & = |2^a\rangle \bar z_a
:= \frac{1}{c}\Big(|\bar{\bm{1}}\rangle + \frac{1}{2m} |q|\bar{\bm{1}}]\Big) , \\
|\bm{2}] & = |2^a] \bar z_a
:= -\frac{1}{c}\Big(|\bar{\bm{1}}] + \frac{1}{2m} |q|\bar{\bm{1}}\rangle \Big) .
\end{aligned}
\end{equation}
Here $|\bar{\bm{1}}\rangle :=  |1^a\rangle \bar{z}_a$ and $|\bar{\bm{1}}] :=  |1^a] \bar{z}_a$ are conjugated states, and the boost-dependent factor $c := \big(1-\tfrac{q^2}{4m^2}\big)^{1/2}$ may be set to unity in the classical limit. 
Note that we have aligned the spin quantization axes of particles~1 and~2, such that the SU(2) wavefunctions coincide, $z_1^a = z^a$ and $z_2^a = \bar{z}^a=(z_a)^*$. This makes the above boost unique.

Then we identify the black-hole spin vector $a^\mu$ with the expectation value of the Pauli-Lubanski spin operator in the spin-$s$ representation, which may be expressed as~\cite{Aoude:2021oqj}
\beal \label{RingRadiusSpinS}
a^\mu = &-(z_a \bar z^a)^{2s-1}\frac{s}{2m^2} \big( \bra{\bar{\bm{1}}}\sigma^\mu|\bm{1}]+\bra{\bm{1}}\sigma^\mu|{\bar{\bm{1}}}] \big) .
\eeal
Furthermore, we set up the classical black-hole spin using a coherent spin state, \ie a superposition of massive spin-$s$ particles.
In terms of the more familiar definite-spin states $\ket{s,\{a\}}$~\cite{Schwinger:1952dse}, it is given by
\begin{equation}
\ket{\text{coherent}} = e^{-z_a\bar{z}^a/2} \sum_{2s=0}^\infty \frac{(z_a)^{\otimes 2s}}{\sqrt{(2s)!}} \ket{s,\{a\}} .
\end{equation}
Here $\{a\} := \{a_1,\dots,a_{2s}\}$ are the symmetrized SU(2) little-group indices (not to be confused with the ring radius vector $a^\mu$) contracted with the variables~$z^a$,
which determine the classical angular momentum~\eqref{RingRadiusSpinS}.

From the Compton amplitude between two massive coherent-spin states and two gravitons, one can extract the classical Compton amplitude
\begin{equation}\label{eq:coherentClAmp}
{\cal M}(\bm{1},\bm{2},3^\pm\!,4^\pm)
 = \lim_{\hbar \to 0} e^{-\bar{z}_a z^a} \!\sum_{2s=0}^\infty \frac{1}{(2s)!} {\cal M}(\bm{1}^s\!,{\bm{2}}^s\!,3^\pm\!,4^\pm) ,
\end{equation}
where the SU(2) wavefunction scale as $z_a\sim \hbar^{-1/2}$ such that $\bar{z}_a z^a \sim \hbar^{-1}$.
In principle, we should also consider off-diagonal terms of the form ${\cal M}(\bm{1}^{s_1}\!,{\bm{2}}^{s_2}\!,3^{\pm}\!,4^{\pm})$ with $s_1\neq s_2$.
However, this would require studying interactions between massive particles of unequal spin, and it is outside the scope of this work.
Here we assume that contributions with $s_1 = s_2$ are sufficiently relevant by themselves in the study of classical Kerr black holes (see \eg~section 3.4 of \rcite{Aoude:2021oqj}), which we corroborate by matching to the classical-gravity literature in the next section. 

To make use of \eqn{eq:coherentClAmp}, we first rewrite the amplitudes~\eqref{eq:quantumcompton} in terms of the classical spin vector, or ring-radius vector, using \eqn{RingRadiusSpinS} and \eqn{eq:boost}.
We use a convenient basis of helicity-independent and dimensionless spin variables,
\beal \label{eq:clComptonVars}
x &: = a \cdot q_\perp , \qquad\;\,\qquad
y := a \cdot q , \\
z & := |a| \frac{p \cdot q_\perp}{m} , \qquad \quad
w := \frac{a \cdot \chi\;p \cdot q_\perp}{p\cdot\chi} ,
\eeal
and the so-called optical parameter $\xi$,
\begin{equation}\label{eq:clGramDet}
\xi^{-1}:=\frac{m^2 q^2}{(p \cdot q_\perp)^2} =\frac{(w - x)^2 - y^2}{z^2-w^2}\,,
\end{equation} 
which is related to the spin variables via a classical-limit Gram determinant.

Using \eqn{eq:coherentClAmp} to resum the finite-spin amplitudes~\eqref{eq:quantumcompton} and including the coupling constant, ${\cal M}_4 = 8 \pi G_\text{N} M_4$, we obtain the following classical amplitude:
\begin{widetext}
\beal
\label{eq:classicalcompton}
{\cal M}(\bm{1},\bm{2},3^-\!,4^+) = {\cal M}_4^{(0)}
\Big(&e^x \cosh z - w \, e^x {\rm sinhc}\, z + \tfrac{w^2 - z^2}{2} E(x,y,z) + (w^2 - z^2) (x-w) \tilde{E}(x,y,z) \\ &
- \frac{(w^2-z^2)^2}{2\xi} \big({\cal E}(x,y,z) +\eta \,\tilde {\cal E}(x,y,z)\big) \Big)+ \alpha\, C_\alpha^{(\infty)} ,
\eeal
\end{widetext}
where we have factored out the spinless Schwarzschild amplitude
${\cal M}_4^{(0)} = - 8\pi G_\text{N} \frac{(p \cdot \chi)^4}{q^2 (p \cdot q_\perp)^2}$. %

The entire functions~$E(x,y,z)$ and $\tilde{E}(x,y,z)$ also appear in the gauge-theoretic root-Kerr amplitudes of~\rcite{Cangemi:2023ysz}, and are defined as
\begin{equation}
E(x,y,z) = \frac{e^y - e^x\!\cosh z + (x{-}y) e^x {\rm sinhc}\,z}{(x-y)^2 - z^2} + (y \to -y)
\end{equation}
and
\beal \label{Etilde}
\tilde{E}(x,y,z) =\!\frac{2 x \cosh y+ (x^2{+}y^2{-}z^2)\,{\rm sinhc}\, y}{\big((x-y)^2 {-} z^2\big)\big((x+y)^2 {-} z^2\big)}+  \big(\substack{y\leftrightarrow z \\ x \to -x} \big) e^x ,
\eeal
where ${\rm sinhc}\, z := z^{-1}\sinh z$ is an even function.

The quantum contact terms specified in \eqn{eq:quantumcompton} generate the last line of the classical amplitude, where the entire functions ${\cal E}$ and $\tilde {\cal E}$ are
\begin{align}
{\cal E}(x,y,z)=& \,
 \frac{e^{x+z}}{2z((x+z)^2-y^2)}  - \frac{e^{x+z}(x+z)}{z((x+z)^2-y^2)^2} 
\nn \\
& +
 \frac{e^{-y}(x+y)}{y ((x+y)^2-z^2)^2}~\,+\, \big(\substack{y \to -y \\ z \to -z}\big),\\
\tilde{\cal E}(x,y,z)= & \frac{e^{x+z}(z-1)}{2z^2((x+z)^2-y^2)} - \frac{e^{x+z}(x+z)}{z((x+z)^2-y^2)^2}\nn \\
& -
  \frac{e^{-y}z}{y ((x+y)^2-z^2)^2}
  ~\,-\, \big(\substack{y \to -y \\ z \to -z}\big) ,
\end{align}
which are derivatives of \eqn{Etilde}:  ${\cal E} = \partial_x \tilde{E}$,  $\tilde{\cal E} = \partial_z \tilde{E}$.
These generate an infinite tower of classical contact terms. Note that the precise combination of the optical parameter, the scalar amplitude ${\cal M}_4^{(0)}$ and the prefactor $(w^2-z^2)^2$ in \eqn{eq:classicalcompton}, conspire to give pole-free terms, see~\rcite{Cangemi:2023ysz}.

Despite their rational form, the four entire functions $E,{\tilde E},{\cal E},{\tilde{\cal E}}$, are analytic everywhere on $\mathbb{C}^3$ and generate an infinite number of spin multipoles. Similar, but simpler, entire functions have been featured in \rcites{Bjerrum-Bohr:2023jau,Bjerrum-Bohr:2023iey,Brandhuber:2023hhl}.

%%%%%%%%%%%%%%%%%%%%%%%%%%%%%%%%%%%%%%%%%%%%%%%%%%%%
\section{Analysis of the Classical Kerr Compton Amplitude}
%%%%%%%%%%%%%%%%%%%%%%%%%%%%%%%%%%%%%%%%%%%%%%%%%%%%

Let us compare our candidate Compton amplitude~\eqref{eq:classicalcompton} to the result of \rcite{Bautista:2022wjf},
where a classical Compton amplitude~${\cal M}_\text{BHPT}(\bm{1},\bm{2},3^-\!,4^+)$ for Kerr black holes was derived, up to $\cO(a^6)$ in the spin-multipole expansion, by studying linear gravitational perturbations of the Kerr spacetime described by the Teukolsky equation.
The final result of \rcite{Bautista:2022wjf} contains terms arising from the expansion of non-analytic functions, such as the digamma function $\psi^{(0)}$, as well as higher polygammas,  which are tagged by a bookkeeping parameter $\alpha$.
Moreover, certain contributions, tagged by $\eta$, are sensitive to the boundary conditions, where $\eta = \pm 1$ corresponds to incoming/outgoing modes at the BH horizon \cite{Bautista:2022wjf}.
Such contributions are therefore interpreted as dissipative effects.

Comparing the classical amplitude~\eqref{eq:classicalcompton} to the Teukolsky result of \rcite{Bautista:2022wjf} up to $\cO(a^6)$, we find perfect agreement up to terms proportional to $\alpha$ after implementing the classical Gram determinant \eqref{eq:clGramDet},
\begin{equation}
\label{eq:teukolskymatch}
{\cal M}(\bm{1},\bm{2},3^-\!,4^+) - {\cal M}_\text{BHPT}(\bm{1},\bm{2},3^-\!,4^+) \Big|_{\alpha=0} = 0 .
\end{equation}
Namely, \eqn{eq:classicalcompton} with $\alpha = 0$ reproduces all the corresponding terms in ${\cal M}_\text{BHPT}$, including the dissipative ones.
\Eqn{eq:teukolskymatch} has been checked up to ${\cal O}(a^6)$ using the results available in \rcite{Bautista:2022wjf}.
Furthermore, we have confirmed the same agreement through $\cO(a^7)$ using the results~\cite{private} shared by the authors of \rcite{Bautista:2022wjf}. Interestingly, this agreement also applies to the terms proportional to the dissipation-related parameter~$\eta$.
We used $\cO(a^5)$ data in order to fix the $\eta$ contact terms in \eqn{eq:quantumcompton} and found that the resulting classical amplitudes correctly predicts $\eta$ terms up to $\cO(a^7)$.
We conclude that the amplitude~\eqref{eq:classicalcompton} predicts the $\alpha = 0$ part of the Kerr Compton amplitude to all orders in spin, although this needs to be confirmed by further results directly coming from general relativity.

In order to capture deviations $\alpha\neq 0$, specifically the ones introduced in \rcite{Bautista:2022wjf} up to sixth order in the multipole expansion, it is instructive to consider the following quantum contact term: 
\def\PB{\mathbb{P}}
\def\QB{\mathbb{Q}}
\begin{equation}\label{ContactAlpha}
C_{\alpha}^{(s)} \!=\!  
  \Big(\!\tfrac{\bra{3}\bar\rho|4]}{2 m^2}+\eta  \tfrac{\bra{3}\rho|4]}{2 m^2}\Big)^{\!3} \Big(\tfrac{\bra{3}\rho|4]}{2}\eta \QB +\tfrac{\bra{3}\bar\rho|4]}{2} ( 3\QB + 2\eta \PB)\!\Big) ,
\end{equation}
where $\bra{3}\bar\rho|4] :=  \braket{3\bm{1}} [\bm{2}4]-\braket{3\bm{2}}[\bm{1}4]$,
and the spin-dependent polynomials are
\beal
\PB &:=\tfrac{1}{m^{4s-8}} \big(P_{5|\vs_1}^{(2s)}- P_{3|\vs_1}^{(2s-2)}  - (\vs_1 \leftrightarrow \vs_2 ) \big), \\
\QB &:= \tfrac{1}{2m^{4s-8}}(\vs_1-\vs_2) \big(P^{(2s)}_{6|\vs_1\vs_1} - P^{(2s)}_{6|\vs_2\vs_2}\big) .
\eeal
When the classical limit is taken, we obtain the identifications: $\PB\rightarrow \tilde {\cal E}(x,y,z)$, $\QB\rightarrow {\cal Q}(x,y,z)$, $\bra{3}\bar\rho|4]\rightarrow 2  w \frac{m^2 p \cdot \chi}{p \cdot q_\perp}$ and $\bra{3}\rho|4]\rightarrow 2 z \frac{m^2 p \cdot \chi}{p \cdot q_\perp}$.
Thus we obtain a simple classical function that captures the $\alpha$-dependent terms of \rcite{Bautista:2022wjf} up to $\cO(a^6)$:
\begin{equation}\label{ClassicalContactAlpha} 
C^{(\infty)}_\alpha 
= -{\cal M}_4^{(0)} \xi^{-1}(w+\eta z)^3 
\big((3w+\eta z)\mathcal{Q} + 2 \eta  w  \tilde {\cal E }
\big) .
\end{equation}
The new entire function is given by derivatives of \eqn{Etilde}, $\mathcal{Q}(x,y,z)=\tfrac{z}{2}\partial_z \partial_x \tilde E$, and the relevant first few orders are 
\begin{equation}
{\cal Q}(x,y,z)= \frac{z^2}{180} + \frac{z^2 x}{252} + {\cal O}(a^4) .
\end{equation}
Beyond ${\cal O}(a^6)$, the simple contact term~\eqref{ContactAlpha} does not suffice for capturing all the $\alpha$-dependent terms, as confirmed through~\rcite{private}. However, some general patterns can be inferred from \eqn{ClassicalContactAlpha} and corresponding ${\cal O}(a^{\ge7})$ generalizations: The $\alpha$-dependent contact terms appear to always be proportional to the spin-dependent factor ${\cal M}_4^{(0)} \xi^{-1}(w+\eta z)^3$, and the remaining dependence is captured by entire functions that vanish in the $z\to 0$ limit. Indeed, all $\eta$- and $\alpha$-dependent terms appear to vanish as $z\to 0$, which corresponds to analytically continuing the spin to a null vector $|a|=0$.
Note, above we have made use of the fact that the parameter $\eta$ is defined on the domain $\eta \in \{1,-1\}$, thus $\eta^2=1$ is an identity.
Also, we consider it illegal to set $\eta=0$ in \eqns{ContactAlpha}{ClassicalContactAlpha}.

It is interesting to study the polar scattering scenario, first considered in~\rcite{Bautista:2022wjf}, where the momentum of the incoming wave is (anti-)aligned with the black-hole spin.
In our notation, this limit corresponds to $w \to \pm z$, and we note that it dramatically simplifies if we also correlate the dissipative parameter $\eta= \mp 1$. Then our classical amplitude \eqref{eq:classicalcompton} behaves as a simple exponential,
\begin{equation}
{\cal M}(\bm{1},\bm{2},3^-\!,4^+)\big|_\text{polar scattering} = {\cal M}_4^{(0)} e^{x\mp z} .
\end{equation}
This exponential behavior is consistent with \rcites{Bautista:2022wjf,private} up to $\cO(a^7)$, and holds to all orders in spin for the classical amplitude proposed in this paper.

%%%%%%%%%%%%%%%%%%%%%%%%%%%%%%%%%%%%%%%%%%%%%%%%%%%%
\section{Discussion}
\label{sec:outro}
%%%%%%%%%%%%%%%%%%%%%%%%%%%%%%%%%%%%%%%%%%%%%%%%%%%%

In this work, we have applied the higher-spin methods we developed in~\rcites{Cangemi:2022bew, Cangemi:2023ysz} to the case of gravitational interactions.
In particular, we have discussed how to construct explicit off-shell cubic higher-spin Lagrangians in the chiral formulation, and used them to reproduce the tree-level three-point amplitude for a Kerr black hole. Moreover, we have used the above Lagrangians to compute the four-point Compton amplitude, allowing for contact term freedom. We found a simple arbitrary-spin expression in terms of the complete homogeneous symmetric polynomials $P_n^{(k)}$, reproducing all known results for spins $s \leq 5/2$ and extending them to higher spins.

We used the coherent-spin framework of~\rcite{Aoude:2021oqj} to compute the classical limit of the quantum amplitude.
We compared it to the Kerr Compton amplitude of \rcite{Bautista:2022wjf}, obtained in BPHT from the Teukolsky equation, and found that our result reproduces all $\alpha=0$ contributions, both conservative and dissipative, computed by the authors up to $\cO(a^7)$ \cite{private}, and extends them to all orders in the black-hole spin vector $a^\mu$.

The treatment of contributions proportional to the book-keeping parameter $\alpha$ remains an open problem. These can be reproduced in our formalism by adding the contact term $C_\alpha^{(\infty)}$  up to $\cO(a^6)$. However, the given contact term is insufficient to predict $\alpha$-dependent contributions at $\cO(a^7)$ and beyond.
It would also be interesting to compare our results to \rcite{Bautista:2023sdf} with $\alpha = 1$, which came out when this letter was in preparation.
This new work handles the polygamma functions differently from \rcite{Bautista:2022wjf}, which brings up the question of whether the related $\alpha$ contributions have an unambiguous definition, and if $\alpha=0$ may be an acceptable tree-level definition. We leave this question to future work.

Our work opens up a number of interesting directions.
Our novel results for the classical Kerr Compton amplitude can be used to extract explicit observables for a binary system of spinning black holes, such as the next-to-leading impulse and spin kick~\cite{Maybee:2019jus}, and the leading-order waveform~\cite{Cristofoli:2021vyo}, to all orders in the black-hole spin and neglecting $\alpha$ contributions.

We can also study the Lagrangians underlying the Kerr Compton amplitude in more detail.
In particular, the dissipative contributions required adding contact terms that break the exchange symmetry between the two massive legs, or alternatively, break the time-reversal symmetry.
This symmetry is guaranteed by a higher-spin Lagrangian in terms of a single field $\Phi^{\alpha_1 \dots \alpha_{2s}}$, therefore breaking it would require introducing additional degrees of freedom, or off-diagonal quantum interactions of different spin. 

The polynomials~$P_n^{(k)}$ played a crucial role in pinpointing gravitational Compton amplitudes, as well as their gauge-theory counterparts in \rcite{Cangemi:2023ysz}.
These naturally arise from imposing parity invariance in the chiral Lagrangians, as well as being a consequence of massive gauge invariance in the non-chiral framework.
It would be interesting to gain a deeper understanding of the connection between these polynomials and fundamental properties of massive higher-spin theory. 

Last but not least, our methods can be extended to five- and higher-point amplitudes, describing non-linear perturbations of Kerr black holes.
This would allow us to compute classical observables at even higher orders, and it would check whether the structures we observe at four points generalize.

\section*{Acknowledgements}

We thank Francesco Alessio, Rafael Aoude, Fabian Bautista, Maor Ben-Shahar, Zvi Bern, Lara Bohnenblust, Gang Chen, Paolo Di Vecchia, Alfredo Guevara, Kays Haddad, Chris Kavanagh, Fei Teng, Radu Roiban, Justin Vines, and Zihan Zhou for helpful discussions.
We are especially grateful to Fabian Bautista, Alfredo Guevara, Chris Kavanagh, and Justin Vines for sharing and discussing their BHPT results up to seventh order in spin.
This research is supported in part by the Knut and Alice Wallenberg Foundation under grants KAW 2018.0116 (From Scattering Amplitudes to Gravitational Waves) and KAW 2018.0162.
The work of M.C. is also supported by the Swedish Research Council under grant 2019-05283.
E.S. is a Research Associate of the Fund for Scientific Research (FNRS), Belgium. The work of E.S. was supported by the European Research Council (ERC) under the European Union’s Horizon 2020 research and innovation programme (grant agreement No 101002551). The work of P.P. was supported by the Science and Technology Facilities Council (STFC) Consolidated Grants ST/T000686/1 “Amplitudes, Strings \& Duality” and ST/X00063X/1 “Amplitudes, Strings \& Duality”. No new data were generated or analyzed during this study.

\bibliographystyle{apsrev4-1}
\bibliography{references}
\end{document}